\documentclass[twocolumn,twoside,slac_two]{revtex4}
\usepackage{graphicx}
\usepackage{fancyhdr}
\def\beq{\begin{equation}}
\def\eeq{\end{equation}}
\def\beqa{\begin{eqnarray}}
\def\eeqa{\end{eqnarray}}

\begin{document}

\title{Higher-order corrections to top-antitop pair and single top quark production}

%

\author{Nikolaos Kidonakis}
\affiliation{Kennesaw State University, Kennesaw, GA 30144, USA}
\begin{abstract}
I present the latest results on the theoretical cross section for top-antitop 
pair production as well as for single top production at the Tevatron and the 
LHC. The calculations include higher-order soft-gluon corrections which are 
dominant near threshold. The top quark transverse
momentum distribution is also presented.
\end{abstract}

\maketitle

\thispagestyle{fancy}


\section{Introduction}

The top quark plays an important role in the Standard Model of particle 
physics as the most massive elementary particle, relevant to 
electroweak physics.
The top was discovered via pair production \cite{CDFttb,D0ttb} 
and has also been more recently 
produced via single top \cite{D0st,CDFst} channels at the Tevatron. 

On the theoretical side, QCD corrections are significant for top 
quark production and even higher-order corrections are not negligible.
An important class of corrections are contributions from soft-gluon 
emission, of the form $\ln^k(s_4/m^2)/s_4$, 
with $s_4$ a variable that measures distance from threshold.
Soft-gluon corrections are dominant near threshold and can be 
resummed. Resummation at NLL (NNLL) accuracy requires one-loop 
(two-loop \cite{NK2l}) calculations in the eikonal approximation. 
Approximate higher-order (NNLO or NNNLO) cross sections can be derived 
from the expansion of the resummed cross section 
\cite{NKRV03,NKNNNLO,NKsttev,NKstlhc,NKRV08}. 

We perform the resummation and higher-order expansions at the differential 
level. In contrast to simple resummation calculations that apply only to 
the total cross section and 
which involve additional approximations for the kinematics, our fully 
differential calculations have sensitivity to the exact kinematics and allow 
not only total cross section calculations but also differential $p_T$ 
and other distributions.

\section{Top pair production}

For top quark hadroproduction the dominant process is 
pair production. At lowest order the partonic processes are 
$q{\bar q}\rightarrow t{\bar t}$ and $gg \rightarrow t{\bar t}$. 
The cross section has been measured at the Tevatron with increasing 
precision \cite{CDFcs,D0cs}. 
There is very good agreement of theory including soft-gluon corrections 
\cite{NKRV08} with Tevatron data. 

\begin{figure}
\begin{center}
\includegraphics[width=8cm]{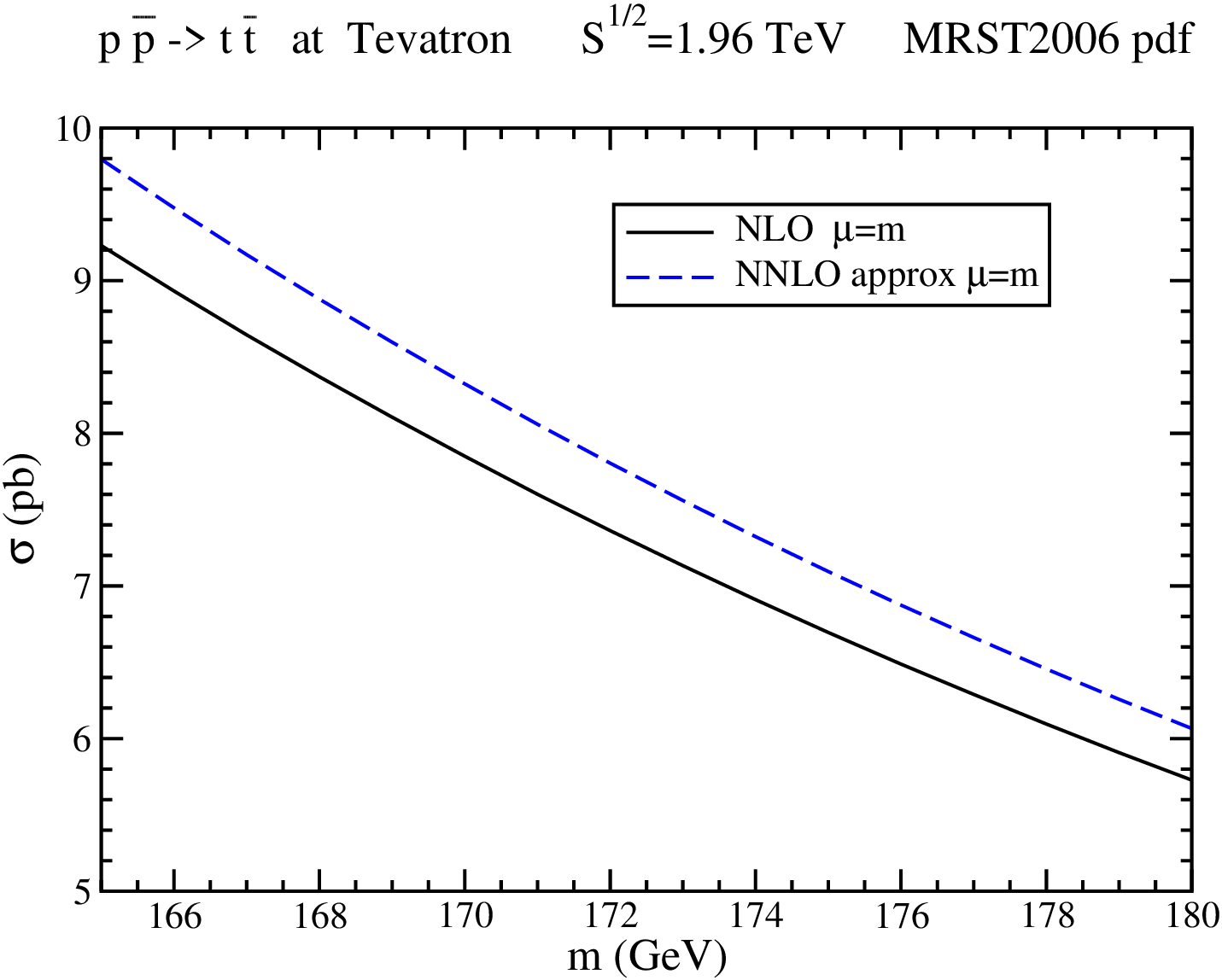}
\caption{The NLO and approximate NNLO $t {\bar t}$ cross sections in 
$p \overline p$ collisions at the Tevatron 
using the MRST 2006 NNLO pdf.}
\end{center}
\end{figure}

\begin{figure}
\begin{center}
\includegraphics[width=8cm]{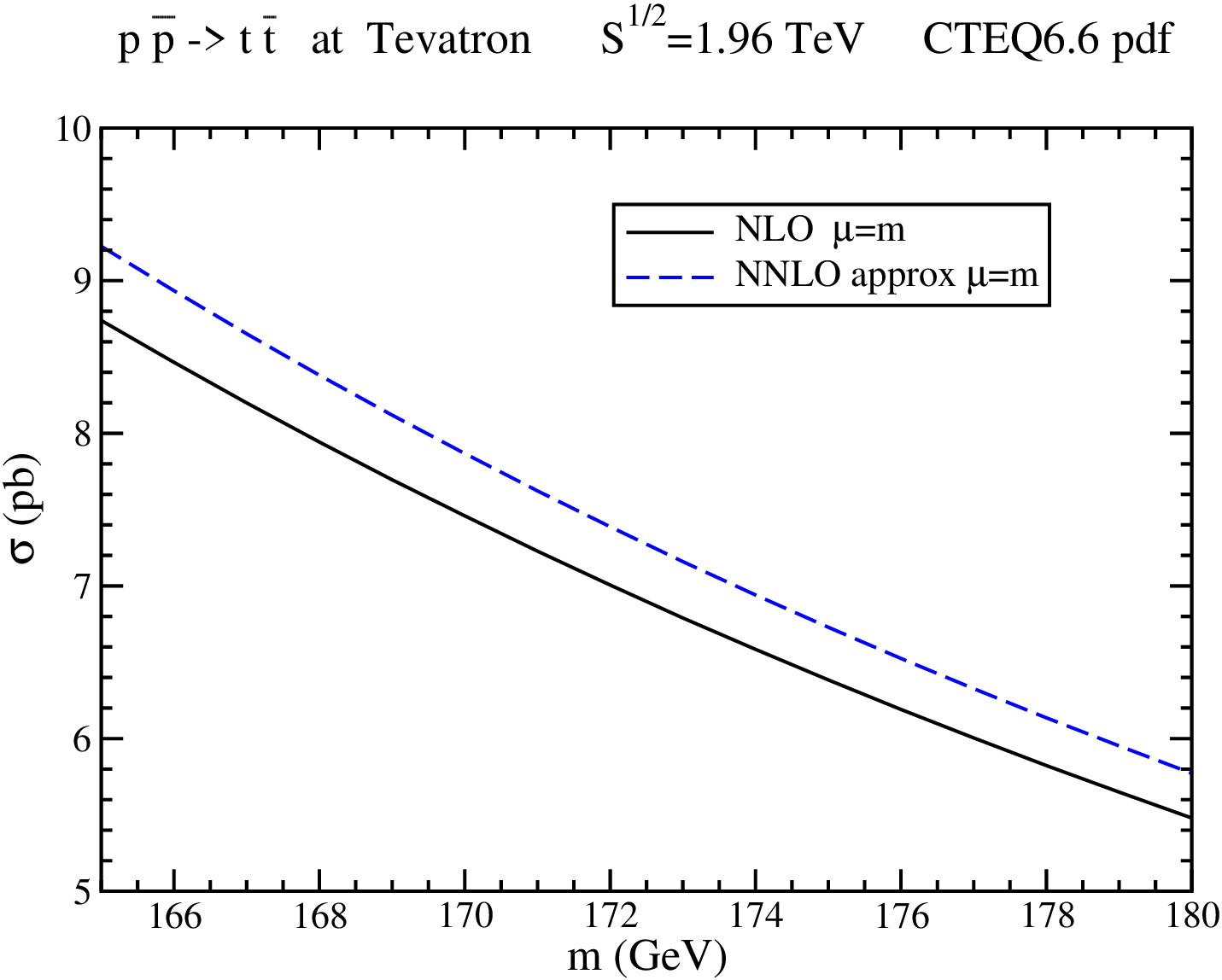}
\caption{The NLO and approximate NNLO $t {\bar t}$ cross sections in 
$p \overline p$ collisions at the Tevatron 
using the CTEQ6.6M pdf.}
\end{center}
\end{figure}

\begin{figure}
\begin{center}
\includegraphics[width=8cm]{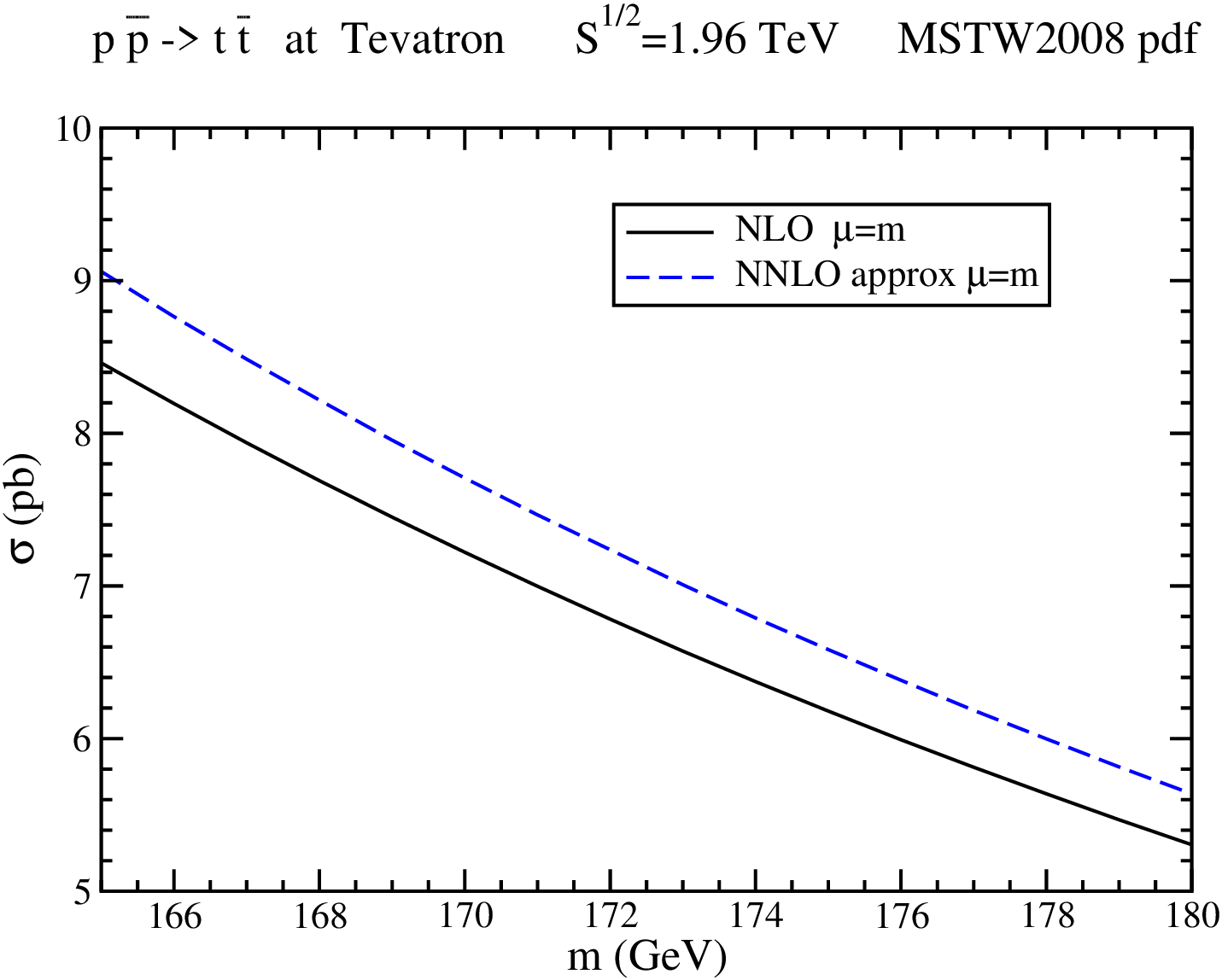}
\caption{The NLO and approximate NNLO $t {\bar t}$ cross sections in 
$p \overline p$ collisions at the Tevatron 
using the MSTW 2008 NNLO pdf.}
\end{center}
\end{figure}

We match at the differential level to the exact NLO \cite{ttbNLO} 
result and add NNLO soft-gluon corrections to obtain approximate NNLO 
cross sections \cite{NKRV03,NKRV08}.
In Fig. 1 we plot the NLO and approximate NNLO cross sections 
for top pair production at the 
Tevatron using the MRST 2006 NNLO pdf \cite{MRST2006}. Figs. 2 and 3 show the corresponding 
results using the CTEQ6.6M \cite{CTEQ6} and MSTW 2008 NNLO pdf \cite{MSTW2008}, respectively. 
The NNLO threshold corrections provide an important enhancement 
as well as a very significant reduction in scale dependence.
Below we give explicit results for top masses of 172 and 173 GeV for all three pdf sets.
For a top quark of 172 GeV the cross sections together with the total uncertainty, calculated 
from the combination of kinematics, scale, and pdf uncertainties, are:
\beqa
\sigma^{\rm NNLOapprox}_{p{\bar p} \rightarrow t \bar t,\, 1.96\,{\rm TeV}}(m=172,{\rm MRST})&=&7.80\; {}^{+0.39}_{-0.45} \; {\rm pb} \, ,
\nonumber \\ 
\sigma^{\rm NNLOapprox}_{p{\bar p} \rightarrow t \bar t,\, 1.96\,{\rm TeV}}(m=172,{\rm CTEQ})&=&7.39\; {}^{+0.57}_{-0.52} \; {\rm pb} \, ,
\nonumber \\ 
\sigma^{\rm NNLOapprox}_{p{\bar p} \rightarrow t \bar t,\, 1.96\,{\rm TeV}}(m=172,{\rm MSTW})&=&7.24\; {}^{+0.30}_{-0.34} \; {\rm pb} \, .
\nonumber 
\eeqa

For a top mass of 173 GeV the corresponding results are shown below: 
\beqa
\sigma^{\rm NNLOapprox}_{p{\bar p} \rightarrow t \bar t,\, 1.96\,{\rm TeV}}(m=173,{\rm MRST})&=&7.56\; {}^{+0.37}_{-0.44} \; {\rm pb} \, ,
\nonumber \\ 
\sigma^{\rm NNLOapprox}_{p{\bar p} \rightarrow t \bar t,\, 1.96\,{\rm TeV}}(m=173,{\rm CTEQ})&=&7.16\; {}^{+0.54}_{-0.50} \; {\rm pb} \, ,
\nonumber \\ 
\sigma^{\rm NNLOapprox}_{p{\bar p} \rightarrow t \bar t,\, 1.96\,{\rm TeV}}(m=173,{\rm MSTW})&=&7.01\; {}^{+0.29}_{-0.33} \; {\rm pb} \, .
\nonumber 
\eeqa
We note that current experimental and theoretical uncertainties are of similar size. Sometimes top masses of 170 and 175 GeV are used in top quark analyses. 
For $m=170$ GeV the result using MSTW pdf is 7.71 pb while for $m=175$ GeV it 
is 6.58 pb. 

\begin{figure}
\begin{center}
\includegraphics[width=8cm]{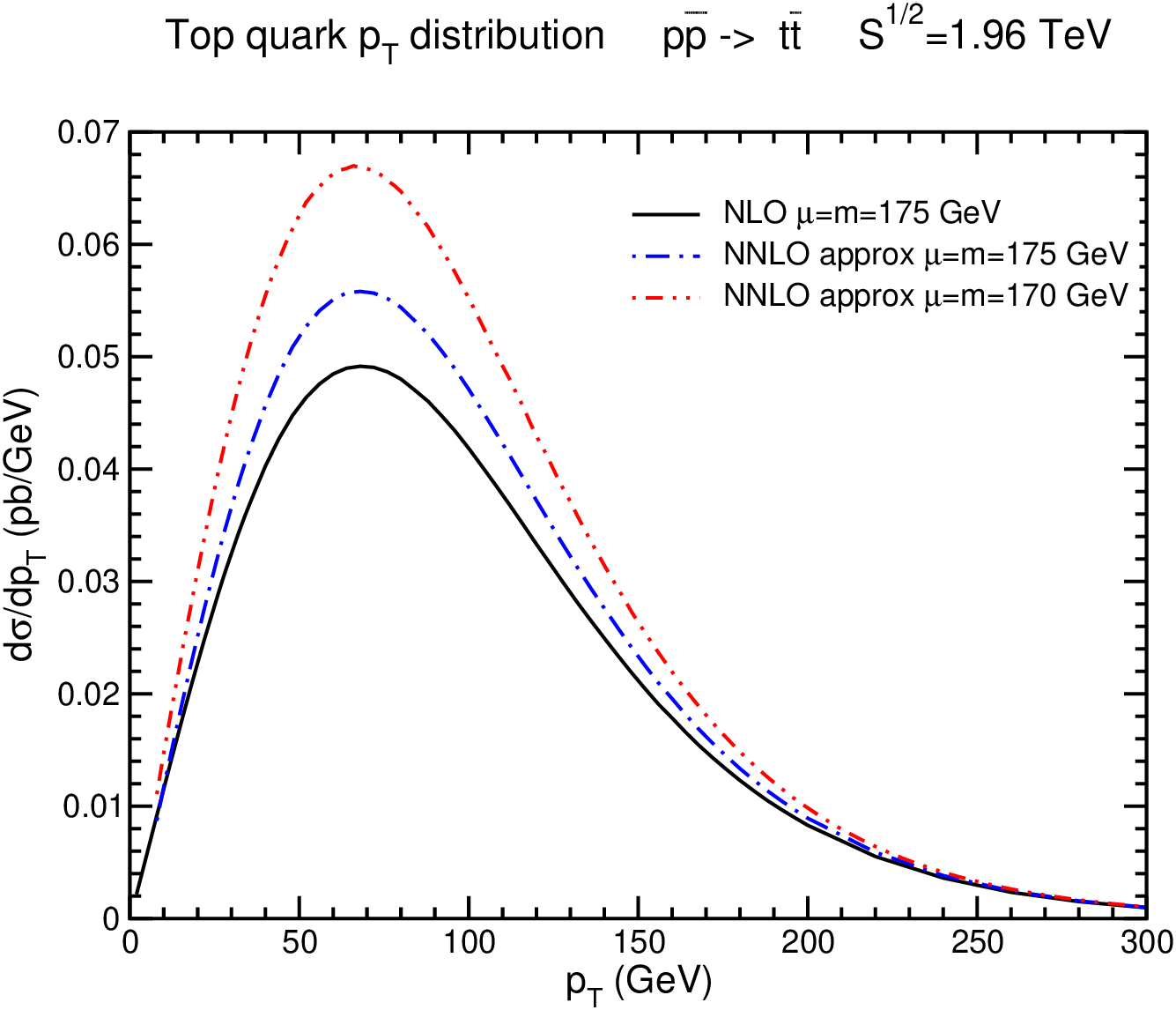}
\caption{The NLO and approximate NNLO top quark tranverse 
momentum distributions in 
$p \overline p$ collisions at the Tevatron.}
\end{center}
\end{figure}

In Fig. 4 we plot the top quark transverse momentum distribution 
at the Tevatron. 
The exact NLO and approximate NNLO results are shown for a top mass of 
175 GeV and, for comparison, also shown is the approximate NNLO result for 
a mass of 170 GeV.
We notice an enhancement at NNLO relative to NLO, but similar shape 
\cite{NKRV03,NKNPA}.

\begin{figure}
\begin{center}
\includegraphics[width=8cm]{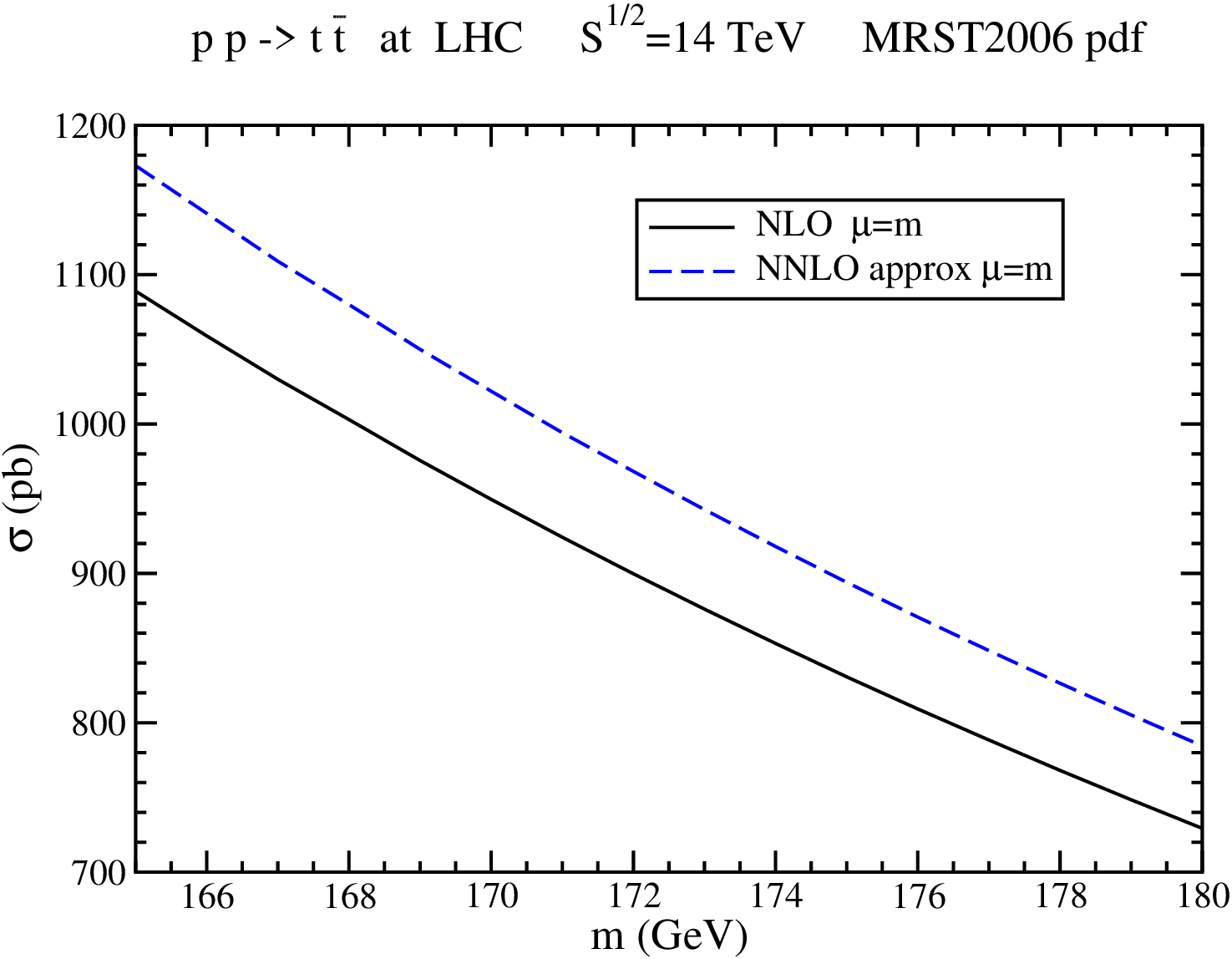}
\caption{The NLO and approximate NNLO $t {\bar t}$ cross sections in 
$p p$ collisions at the LHC at 14 TeV  
using the MRST 2006 NNLO pdf.}
\end{center}
\end{figure}

\begin{figure}
\begin{center}
\includegraphics[width=8cm]{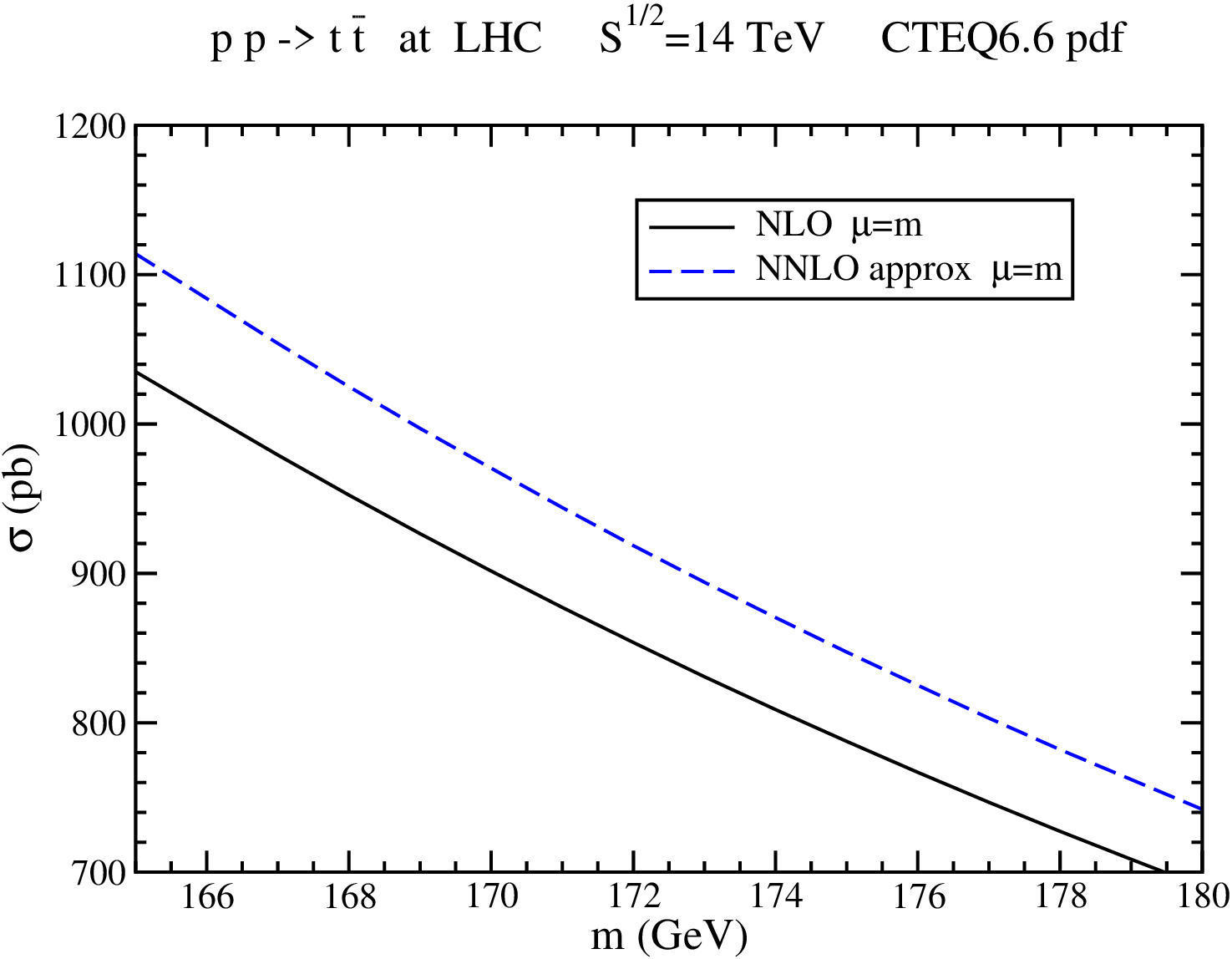}
\caption{The NLO and approximate NNLO $t {\bar t}$ cross sections in 
$p p$ collisions at the LHC at 14 TeV
using the CTEQ6.6M pdf.}
\end{center}
\end{figure}

\begin{figure}
\begin{center}
\includegraphics[width=8cm]{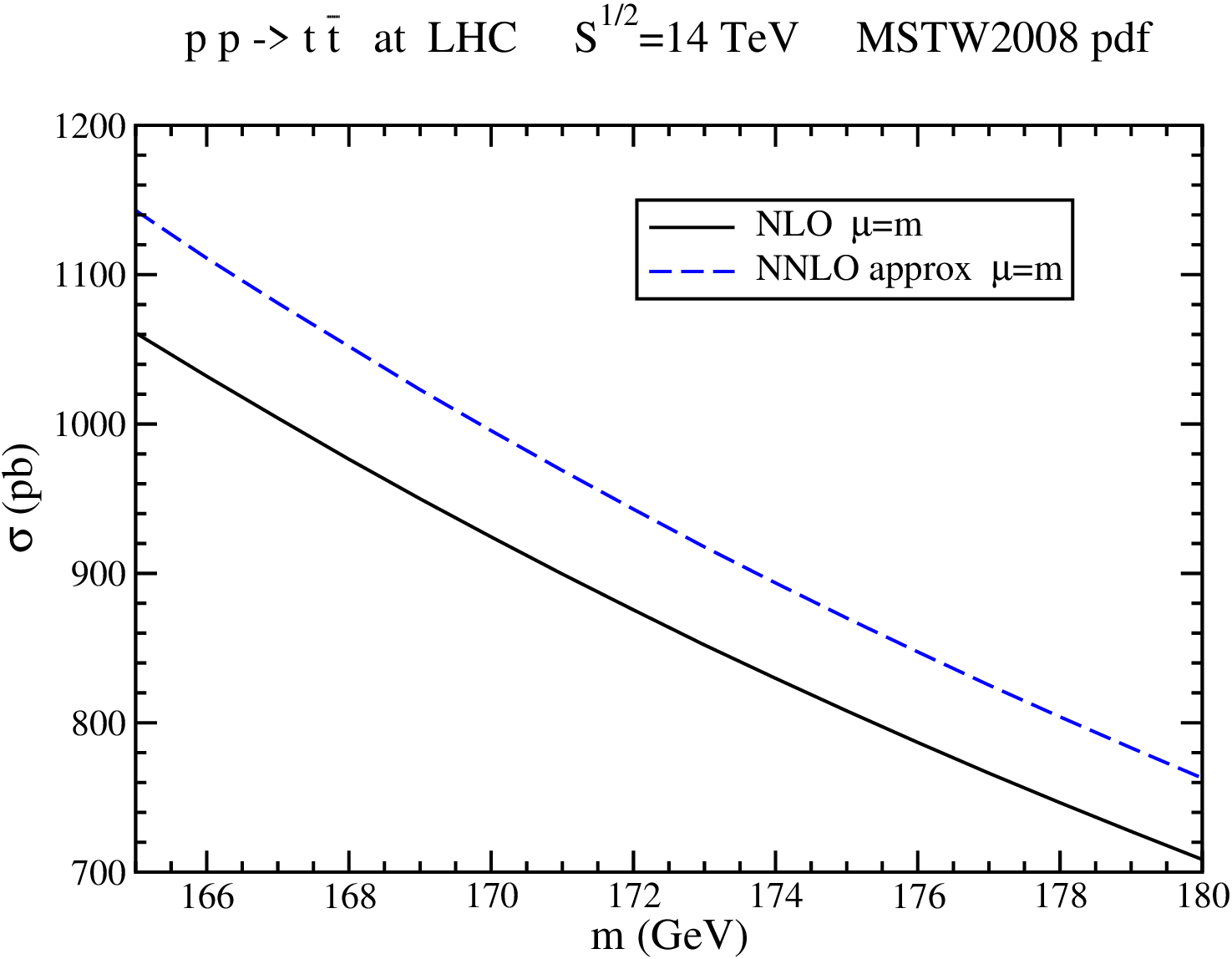}
\caption{The NLO and approximate NNLO $t {\bar t}$ cross sections in 
$p p$ collisions at the LHC at 14 TeV
using the MSTW 2008 NNLO pdf.}
\end{center}
\end{figure}

Next, we discuss the top quark pair cross section at the LHC.
In Fig. 5 we plot the NLO and approximate NNLO cross sections  
for top pair production at the LHC at 14 TeV 
using the MRST 2006 NNLO pdf. Figs. 6 and 7 show, respectively, the results using the CTEQ6.6M and MSTW 2008 NNLO pdf. Again, the NNLO threshold corrections provide an overall enhancement and reduction in scale dependence.
For a top quark of 172 GeV the cross sections together with the total uncertainty are
\beqa
\sigma^{\rm NNLOapprox}_{pp \rightarrow t \bar t,\, 14\,{\rm TeV}}(m=172,{\rm MRST})&=&968 \; ^{+80}_{-52} \; {\rm pb} \, ,
\nonumber \\ 
\sigma^{\rm NNLOapprox}_{pp \rightarrow t \bar t,\, 14\,{\rm TeV}}(m=172,{\rm CTEQ})&=&919 \; {}^{+76}_{-55} \; {\rm pb} \, ,
\nonumber \\ 
\sigma^{\rm NNLOapprox}_{pp \rightarrow t \bar t,\, 14\,{\rm TeV}}(m=172,{\rm MSTW})&=&943 \; ^{+66}_{-42} \; {\rm pb} \, .
\nonumber
\eeqa
For a top mass of 173 GeV the corresponding results are
\beqa
\sigma^{\rm NNLOapprox}_{pp \rightarrow t \bar t, \, 14\,{\rm TeV}}(m=173,{\rm MRST})&=&943 \; ^{+78}_{-51} \; {\rm pb} \, ,
\nonumber \\ 
\sigma^{\rm NNLOapprox}_{pp \rightarrow t \bar t, \, 14\,{\rm TeV}}(m=173,{\rm CTEQ})&=&894 \; {}^{+74}_{-54} \; {\rm pb} \, ,
\nonumber \\ 
\sigma^{\rm NNLOapprox}_{pp \rightarrow t \bar t, \, 14\,{\rm TeV}}(m=173,{\rm MSTW})&=&918 \; ^{+64}_{-41} \; {\rm pb} \, .
\nonumber
\eeqa

\begin{figure}
\begin{center}
\includegraphics[width=8cm]{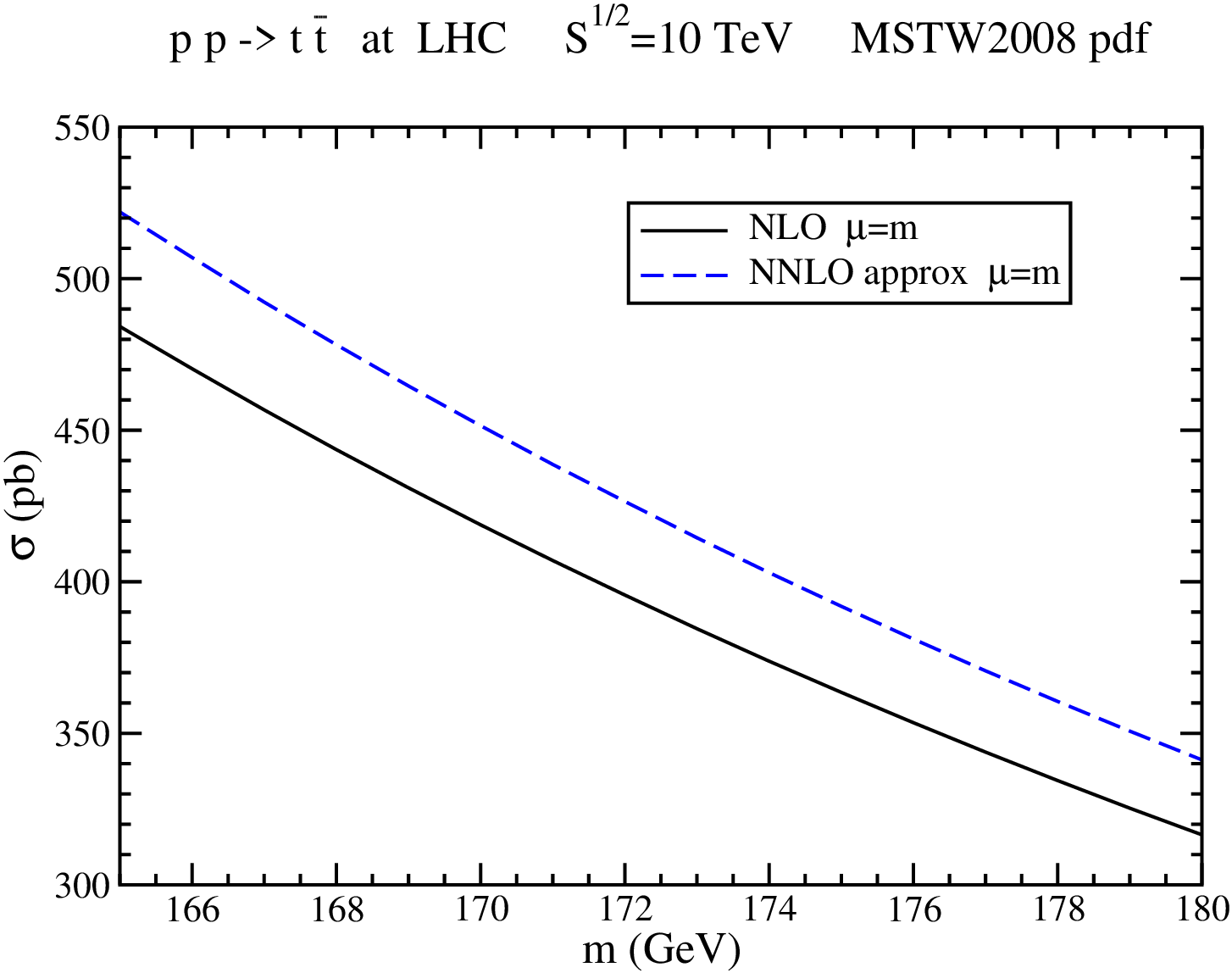}
\caption{The NLO and approximate NNLO $t {\bar t}$ cross sections in 
$p p$ collisions at the LHC at 10 TeV
using the MSTW 2008 NNLO pdf.}
\end{center}
\end{figure}

In Fig. 8 the top pair cross section is plotted for an LHC energy of 10 TeV.
At 10 TeV, using the MSTW pdf, the cross section is 
\beqa
\sigma^{\rm NNLOapprox}_{pp \rightarrow t \bar t,\, 10\,{\rm TeV}}(m=172, {\rm MSTW})
&=&427\; ^{+32}_{-19} \; {\rm pb} \, ,
\nonumber \\
\sigma^{\rm NNLOapprox}_{pp \rightarrow t \bar t,\, 10\,{\rm TeV}}(m=173, {\rm MSTW})
&=&415\; ^{+31}_{-19} \; {\rm pb} \, .
\nonumber
\eeqa

\begin{figure}
\begin{center}
\includegraphics[width=8cm]{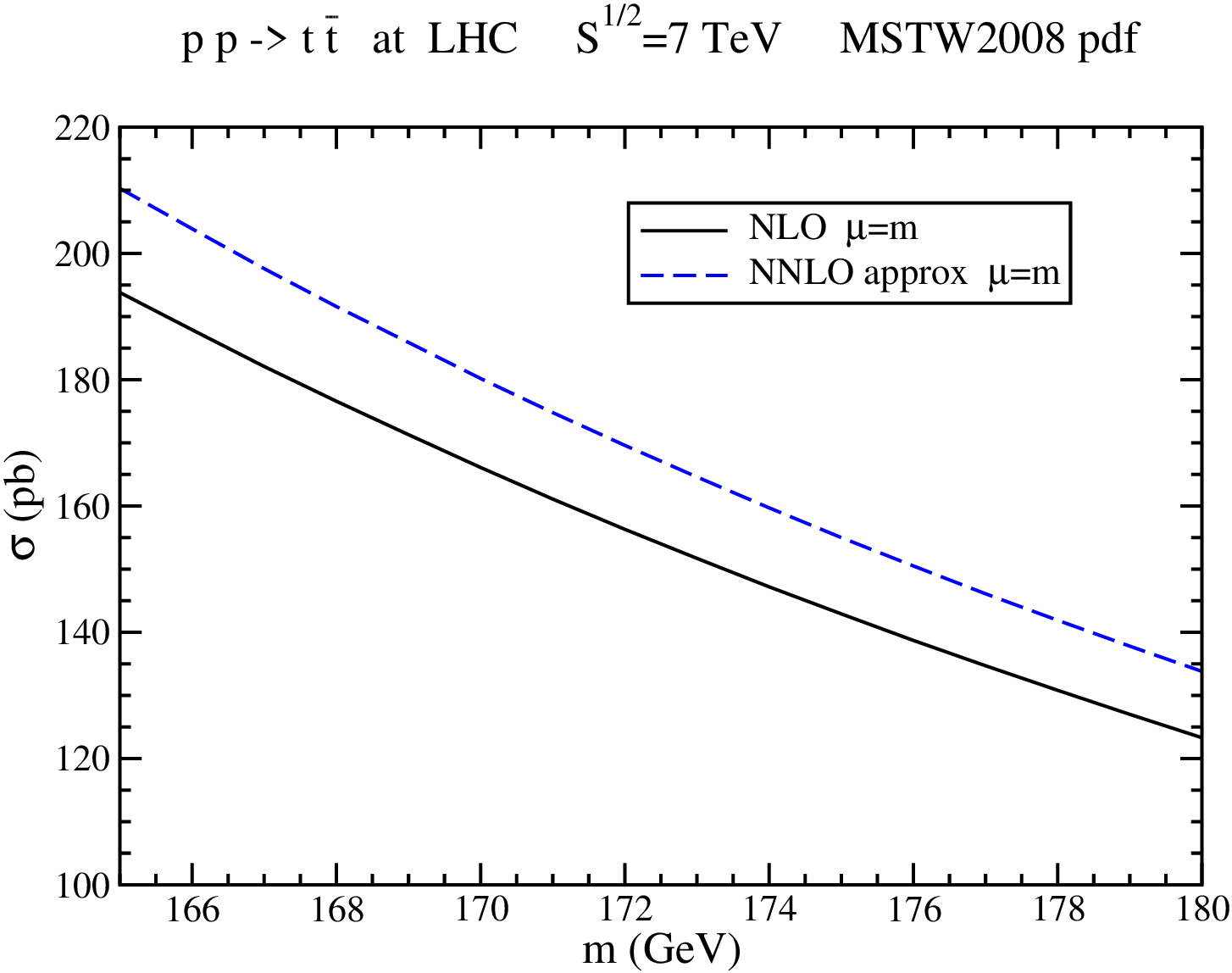}
\caption{The NLO and approximate NNLO $t {\bar t}$ cross sections in 
$p p$ collisions at the LHC at 7 TeV
using the MSTW 2008 NNLO pdf.}
\end{center}
\end{figure}

In Fig. 9 the $t{\bar t}$ cross section is shown for the starting LHC energy of 7 TeV. 
At 7 TeV, using the MSTW pdf, we find 
\beqa
\sigma^{\rm NNLOapprox}_{pp \rightarrow t \bar t,\, 7\,{\rm TeV}}(m=172, {\rm MSTW})
&=&170 \pm 10 \; {\rm pb} \, ,
\nonumber \\
\sigma^{\rm NNLOapprox}_{pp \rightarrow t \bar t,\, 7\,{\rm TeV}}(m=173, {\rm MSTW})
&=&165 \pm 10 \; {\rm pb} \, .
\nonumber
\eeqa

The top quark transverse momentum distribution at the LHC is enhanced 
at NNLO relative to NLO, but with little change in shape \cite{NKRV03,NKNPA}.

\section{Single top quark production}

The recent observation of single top production \cite{D0st,CDFst} provides
opportunities for the study of electroweak properties of the top, 
such as extracting the $V_{tb}$ CKM matrix element \cite{sintopcombo}.
Single top production can proceed via three partonic channels:
the $t$-channel, $qb \rightarrow q' t$ and ${\bar q} b \rightarrow {\bar q}' t$; 
the $s$-channel, $q{\bar q}' \rightarrow {\bar b} t$; and
associated $tW$ production, $bg \rightarrow tW^-$.

\begin{figure}
\begin{center}
\includegraphics[width=8cm]{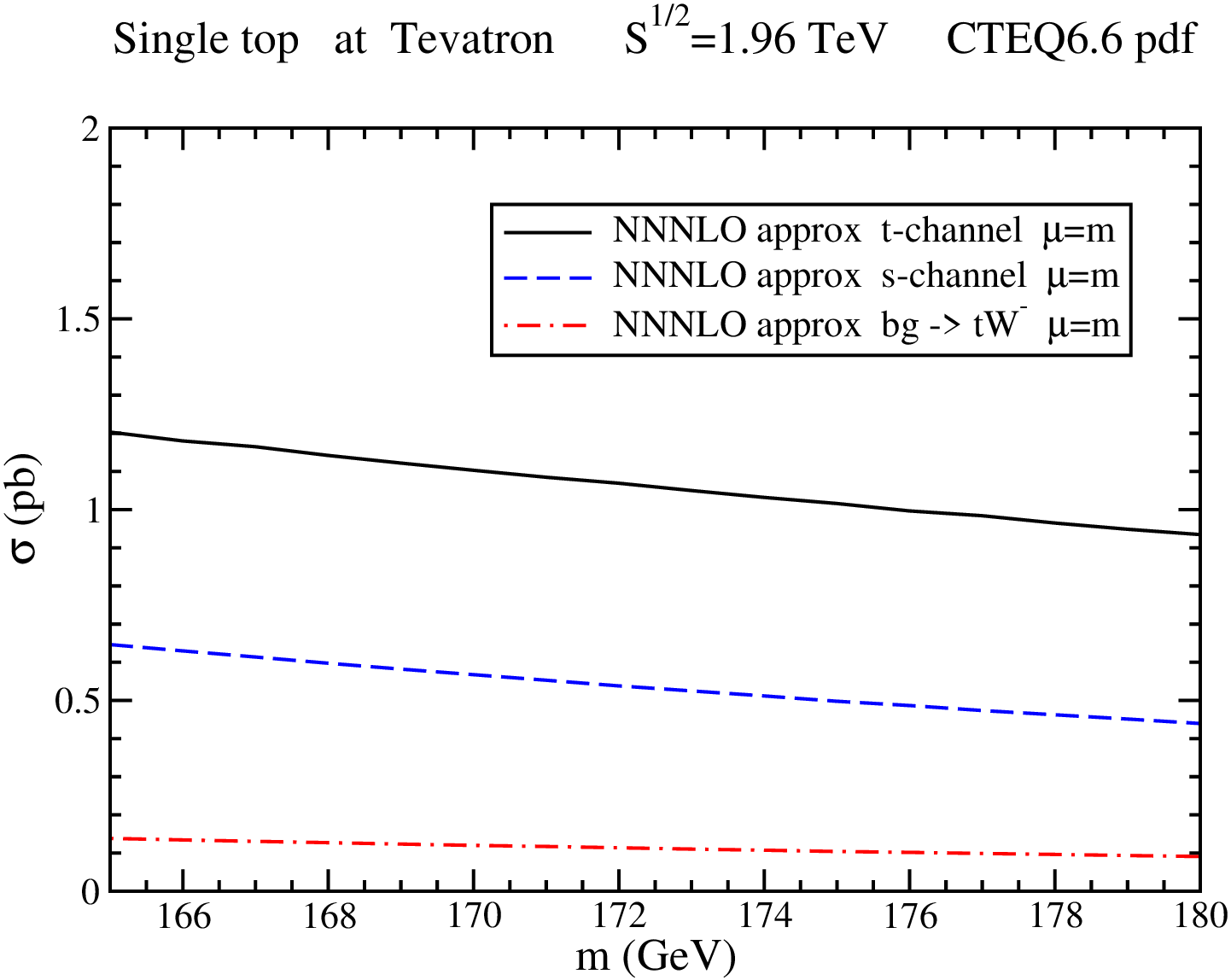}
\caption{The approximate NNNLO single top cross sections in 
the $t$, $s$, and $bg\rightarrow tW^-$ channels at the Tevatron 
using the CTEQ6.6M pdf.}
\end{center}
\end{figure}

Figure 10 shows the approximate NNNLO cross sections for single top production 
at the Tevatron in all three channels using CTEQ6.6M pdf
(note that the cross sections for anti-top production are the same).
The $t$-channel has the largest cross section followed by the $s$-channel, 
while $bg\rightarrow tW^-$ is relatively small. 
Below are listed the cross sections for $m=172$ GeV: 
\beqa
\sigma^{\rm NNNLOapprox}_{t-{\rm chan},\, 1.96\,{\rm TeV}}(m=172, \rm CTEQ)&=&1.07 \pm 0.11 \; {\rm pb} \, ,
\nonumber \\
\sigma^{\rm NNNLOapprox}_{s-{\rm chan},\, 1.96\,{\rm TeV}}(m=172, \rm CTEQ)&=&0.54 \pm 0.03 \; {\rm pb} \, ,
\nonumber \\
\sigma^{\rm NNNLOapprox}_{bg \rightarrow tW,\, 1.96\,{\rm TeV}}(m=172, \rm CTEQ)&=&0.11 \pm 0.04 \; {\rm pb} \, ,
\nonumber
\eeqa
and for $m=173$ GeV:
\beqa
\sigma^{\rm NNNLOapprox}_{t-{\rm chan},\, 1.96\,{\rm TeV}}(m=173, \rm CTEQ)&=&1.05 \pm 0.11 \; {\rm pb} \, ,
\nonumber \\
\sigma^{\rm NNNLOapprox}_{s-{\rm chan},\, 1.96\,{\rm TeV}}(m=173, \rm CTEQ)&=&0.52 \pm 0.03 \; {\rm pb} \, ,
\nonumber \\
\sigma^{\rm NNNLOapprox}_{bg\rightarrow tW,\, 1.96\,{\rm TeV}}(m=173, \rm CTEQ)&=&0.11 \pm 0.04 \; {\rm pb} \, .
\nonumber
\eeqa
Detailed results using MRST pdf are given in \cite{NKsttev,NKNPA,NKAPP}.

At the LHC the $t$-channel is numerically dominant and $tW$ production 
is second in cross section. The $s$-channel, which is second largest at the 
Tevatron, is relatively small at the LHC.

At LHC energies the threshold approximation does not work well for the 
$t$-channel so only NLO results are known, but approximate NNNLO results 
for the $s$-channel and $tW$ production are 
given in \cite{NKstlhc,NKNPA,NKAPP}. 
More work on NNLL soft-gluon corrections 
and on purely collinear corrections \cite{NKcol} 
may improve the approximation for LHC energies.

\begin{acknowledgments}
This work was supported by the National Science Foundation under 
Grant No. PHY 0855421.
\end{acknowledgments}

\bigskip 

\end{document}